\documentclass[a4paper,11pt]{article}
\usepackage{graphicx} 
\usepackage{subcaption}

\pdfoutput=1 

\usepackage{jcappub} 

\usepackage{hanging}
\usepackage{orcidlink}
\usepackage[T1]{fontenc} 
\usepackage{multirow} 

\usepackage{pifont}

\newcommand{\bq}{\boldsymbol q}

\newcommand{\bk}{\boldsymbol k}

\newcommand{\ihmpc}{\,h{\rm Mpc}^{-1}}

\title{An analytic approximation to the covariance between pre- and post-reconstruction galaxy two-point statistics}
\author[1,2,3]{{M.~Maus}\orcidlink{0000-0002-9020-911X},}
\author[1,2,3]{{A.~Baleato Lizancos}\orcidlink{0000-0002-0232-6480},}
\author[1,2,3]{{M.~White}\orcidlink{0000-0001-9912-5070},}
\author[4]{{A.~de Mattia}\orcidlink{0000-0003-0920-2947}}
\author[5,6]{{and S.~Chen}\orcidlink{0000-0002-5762-6405}}

\affiliation[1]{Berkeley Center for Cosmological Physics, UC Berkeley, CA 94720, USA}
\affiliation[2]{Department of Physics, University of California, Berkeley, CA 94720, USA}
\affiliation[3]{Lawrence Berkeley National Laboratory, One Cyclotron Road, Berkeley, CA 94720, USA}
\affiliation[4]{IRFU, CEA, Université Paris-Saclay, F-91191 Gif-sur-Yvette, France}
\affiliation[5]{Department of Physics, Columbia University, New York, NY 10027, USA}
\affiliation[6]{NASA Hubble Fellowship Program, Einstein Fellow}

\emailAdd{mark.maus@berkeley.edu}
\emailAdd{a.baleatolizancos@berkeley.edu}
\emailAdd{mwhite@berkeley.edu}

\abstract{
We present a simple analytic approximation for the covariance between pre-reconstruction galaxy power spectrum measurements and post-reconstruction two-point correlation functions. This cross-covariance is essential for joint analyses that combine full-shape clustering information with baryon acoustic oscillation (BAO) measurements, as commonly performed in modern spectroscopic surveys. Our model builds on the disconnected contribution to the covariance and accounts for the damping of correlations due to the BAO reconstruction process. We validate our analytic prescription against numerical simulations from the Dark Energy Spectroscopic Instrument (DESI), testing both idealized cubic geometries and realistic survey configurations including complex footprints and fiber assignment effects. Despite neglecting survey window functions in the analytic calculation, we find excellent agreement with simulation-based covariances and demonstrate that cosmological parameter constraints are virtually unchanged when using our approximation. Our results show that the pre-post cross-covariance is sufficiently small that even approximate treatments are adequate for cosmological inference, opening a pathway toward fully analytic covariance matrices for next-generation galaxy surveys.
}

\begin{document}

\maketitle

\section{Introduction}

Few observables have taught cosmologists more about the constituents of the Universe and its large-scale properties than the positions of galaxies have. Modern spectroscopic surveys like DESI~\cite{collaborationDESIExperimentPart2016}, Euclid~\cite{Euclid}, PFS~\cite{sugaiPrimeFocusSpectrograph2015} or 4MOST~\cite{4MOST4metreMultiObject} are able to measure precise angular positions and redshifts for millions of objects. When analyzed statistically, the clustering pattern they encode can be compared to predictions from theoretical models and used to validate or refute them.

Most readily apparent in the clustering pattern of galaxies is the relative excess clustering seen on scales corresponding to the comoving size of the sound horizon at the epoch of photon-baryon decoupling -- the left over imprint of the baryon acoustic oscillations (BAO). Since gravitational evolution tends to smear out this valuable observational feature, `reconstruction' techniques have been developed to partially undo the smearing effect and sharpen probes of the BAO feature~\cite{eisensteinImprovingCosmologicalDistance2007}. It can be shown that BAO reconstruction works by retrieving information from higher-order statistics and making it accessible from lower-order ones like the power spectrum or the two-point correlation function (2PCF). The latest in a long lineage of redshift surveys, DESI has recently used this technology to chart the expansion history of the Universe and place tight constraints on the cosmological model~\cite{DESI_DR2_BAO}.

In parallel, with the increase in statistical power enabled by ever growing volumes of galaxy survey data, a greater emphasis has been placed on extracting information beyond the BAO feature. This is motivated by the fact that the `full shape' of the clustering statistics responds to the energy contents of the Universe, the spatial geometry of the background, and the properties of gravity. In recent years, such full-shape analyses have become a mature and competitive cosmological probe (see e.g.~\cite{collaborationDESI2024VII2024,Maus25, novell-masotFullShapeAnalysisPower2025, chudaykinReanalyzingDESIDR12025a, chudaykinReanalyzingDESIDR12026, ivanovReanalyzingDESIDR12026} for recent analyses of DESI data).

Several modern analyses have jointly analyzed the full-shape of galaxy clustering statistics alongside the BAO feature post-reconstruction~\cite{collaborationDESI2024VII2024, Chen22, Chen24, Maus25, chudaykinReanalyzingDESIDR12025a, chudaykinReanalyzingDESIDR12026, ivanovReanalyzingDESIDR12026}, seeking to extract the complementary insights they encode. However, these two observables also share some amount of common information, which means that any rigorous analysis of this kind must account for the covariance between them. 

There are in principle several ways to estimate this covariance. The first is by looking at the scatter of the statistics measured on subsets of the data --- e.g. bootstrap or jackknife techniques, as described in~\cite{philcoxRascalCJackknifeApproach2020, KP4s7-Rashkovetskyi} and references therein. However, the usefulness of these techniques on the large scales where the cosmological information resides is limited by cosmic variance -- the fact that we can only build a small number of large-scale data splits. The leading alternative is to resort to simulations. This bypasses the aforementioned limitation while still lending itself to rather easily incorporating observational effects such as incomplete sky coverage, fiber collisions and other effects which collectively constitute the ``observational window''. Simulations in principle also allow us to track the impact of non-linear physics, though this can in fact turn out to be a weakness when the underlying model is inaccurate, potentially leading to biased inference~\cite{baumgartenRobustnessCovarianceMatrix2018a}. More importantly, numerical estimates of the covariance are very costly, typically requiring thousands of runs that can only be completed on the most powerful of supercomputers~\cite{taylorPuttingPrecisionPrecision2013}.

The final route -- an analytic calculation -- thus remains very appealing, as it would in principle resolve the aforementioned issues by allowing much greater flexibility and computational efficiency. Moreover, analytic covariance matrices are noiseless, and can even be used to enhance the numerical route by denoising the simulation-based covariance. Though the connected part of the covariance is likely too complicated to be treated analytically, being sensitive to baryonic effects, beat-coupling, super-sample covariance and shot noise (see~\cite{wadekarGalaxyPowerSpectrum2020} and references therein), previous work has demonstrated that on large scales the full covariance matrix is well approximated by an analytic calculation of the disconnected component, which even offers a viable route to include window effects~\cite{Li2019,KP4s6-Forero-Sanchez}.

But perhaps owing to the perception that the reconstruction process and its interplay with the ``observation window'' would complicate such an analytic treatment, to date there have been no efforts to describe analytically the covariance between the galaxy 2PCF post-reconstruction and the pre-reconstruction power spectrum. In this work, we provide a very simple prescription to do this, and argue that the correlation is small enough that even an approximate treatment is sufficient.

Recently, ref.~\cite{chudaykinReanalyzingDESIDR12026} used an analytic approximation to the pre- and post-reconstruction power spectrum covariance calibrated to measured statistics. Here, we choose to work with the post-reconstruction 2PCF rather than the power spectrum. Since the BAO information is confined to a very narrow range of scales in this statistic \cite{KP4s2-Chen}, the correlation between the post-reconstruction 2PCF and pre-reconstruction power spectrum is small and comes strictly from the physics of BAO. Our analytic model \emph{predicts} this correlation very accurately given a set of cosmological and galaxy bias parameters. Such a fully analytic prediction would be more complicated if we were working instead with compressed statistics relying on fixed templates like the BAO parameters $\alpha_\parallel$ and $\alpha_\perp$, although our calculation may provide a pathway to calculating this; see e.g.~\cite{Chen22,Maus23, KP4s2-Chen,KP5s2-Maus,Maus25} for other reasons why the direct modeling approach is to be preferred over compressed statistics.

This paper is structured as follows. In \S\ref{sec:model}, we provide our simple, analytic model for the covariance between the pre-reconstruction power spectrum and the post-reconstruction 2PCF. Then, in \S\ref{sec:sims}, we validate the accuracy of our approximation against numerical simulations. In appendix~\ref{sec:window_rot}, we investigate how rotation of the window matrix in the context of fiber assignment mitigation impacts our results. Finally, \S\ref{sec:conclusions} presents our conclusions.

\section{Analytic model}\label{sec:model}

Assuming that the cross-correlation between the pre- and post-reconstruction clustering statistics can be reasonably well approximated by its disconnected (or ``Gaussian'') contribution, it is straightforward to derive the covariance between the Legendre multipoles of the pre-reconstruction power spectrum, $P_{\ell}^{\rm pre}(k)$, and the post-reconstruction 2PCF, $\xi_\ell^{\rm post}(r)$, if we neglect the survey ``window''.  First note that the power spectrum Legendre multipoles are simple integrals of the anisotropic $P^{\rm pre}$ over angles and $k$-bins,
\begin{equation}
    P_\ell^{\rm pre}(k) = (2\ell+1)\int\frac{d^3k}{4\pi\,k^2\Delta k}
    P^{\rm pre}(\mathbf{k})\mathcal{L}_\ell\left(\widehat{k}\cdot\widehat{z}\right)\,,
\end{equation}
where $\mathcal{L}_\ell$ is the Legendre polynomial of order $\ell$, $\widehat{z}$ represents the line-of-sight direction and we shall assume $P^{\rm pre}$ varies slowly over the $k$ bin of width $\Delta k$.  The (post-reconstruction) correlation function multipoles are
\begin{equation}
    \xi_\ell^{\rm post}(r) = (2\ell+1) i^\ell
    \int \frac{d^3k}{(2\pi)^3} P^{\rm post}(\mathbf{k})
    \mathcal{L}_\ell\left(\widehat{k}\cdot\widehat{z}\right)
    j_\ell(kr)\,,
\end{equation}
where $j_\ell$ is the spherical Bessel function of order $\ell$.
We can account for averaging $\xi_\ell$ over a bin of width $\Delta r$ by replacing $j_\ell\to\bar{j}_\ell$ with the obvious definition of $\bar{j}_\ell$.
The pre-post-covariance is then
\begin{align}
    \mathrm{Cov}\left[ P_{\ell_1}^{\rm pre}(k),\xi_{\ell_2}^{\rm post}(r) \right]
    &= (2\ell_1+1)(2\ell_2+1)\, i^{\ell_2} \int\frac{d^3k_1}{4\pi\,k^2\Delta k}
    \int \frac{d^3k_2}{(2\pi)^3} 
    \nonumber \\
    &\times 
    \mathcal{L}_{\ell_1}\left(\widehat{k}_1\cdot\widehat{z}\right)
    \mathcal{L}_{\ell_2}\left(\widehat{k}_2\cdot\widehat{z}\right)
    \bar{j}_{\ell_2}(k_2r) \nonumber \\
    &\times
    \mathrm{Cov}\left[ P^{\rm pre}(\mathbf{k}_1) , P^{\rm post}(\mathbf{k}_2) \right]\,.
\end{align}
If the covariance is dominated by the disconnected contribution then
\begin{equation}
    \mathrm{Cov}\left[ P^{\rm pre}(\mathbf{k}_1) , P^{\rm post}(\mathbf{k}_2) \right]
    = P_{\times}^2(\mathbf{k})\ \frac{(2\pi)^3}{V}
    \left[ \delta^{(D)}(\mathbf{k}_1+\mathbf{k}_2)+\delta^{(D)}(\mathbf{k}_1-\mathbf{k}_2)\right]
\end{equation}
where $V$ is the survey volume we have used $P_\times$ for the cross-power between the pre- and post-reconstruction density fields.  Assuming $\ell_1$ and $\ell_2$ are even, the two contributions are equal and
\begin{align}
    \mathrm{Cov}\left[ P_{\ell_1}^{\rm pre}(k),\xi_{\ell_2}^{\rm post}(r) \right]
    &= \frac{(2\ell_1+1)(2\ell_2+1)}{V} \bar{j}_{\ell_2}(kr)
    \sum_{L_1L_2} P_{\times,L_1}(k) P_{\times,L_2}(k) \nonumber \\
    &\times  i^{\ell_2} \int d\mu
    \ \mathcal{L}_{\ell_1}(\mu) \mathcal{L}_{\ell_2}(\mu)
    \,\mathcal{L}_{L_1}(\mu)\mathcal{L}_{L_2}(\mu)
    \quad .
    \label{eq:cov}
\end{align}
If we were to assume that $P_{\times,0}$ dominates over all other $L$ then $L_1=L_2=0$ and thus $\ell_1$ must equal $\ell_2$.  For example for $\ell_1=\ell_2=0$ we get $(2/V) \bar{j}_0 P_{\times,0}^2$ while for $\ell_1=\ell_2=2$ it is $(-10/V)\bar{j}_2 P_{\times,0}^2$.  This makes a useful limit for cross-checks.

To complete the model we need an approximation for $P_\times$.  Within perturbation theory and on large scales \cite{Sugiyama24,KP4s2-Chen}
\begin{equation}
    P_{\times} \approx e^{-k^2 \sigma^2(\mu)/2}
    \left[ \left(b+f\mu^2\right)^2\, P_{\rm lin}(k) + \mathrm{SN} \right]\,,
    \label{eq:Px}
\end{equation}
where $b$ is the large-scale bias, SN is the shot-noise/stochastic contribution and
\begin{equation}
    \sigma^2(\mu) \approx \left[1+(1+f)^2\mu^2\right]
    \ \frac{1}{3} \int \frac{dk}{2\pi^2}\ P_{\rm lin}(k)\,W^2(k)\,,
    \label{eq:sig}
\end{equation}
with $W$ the window function of the (typically Gaussian) smoothing applied when computing the displacements in the process of reconstruction.  Further details on the approximations inherent in Equations \ref{eq:Px} and \ref{eq:sig} can be found in Appendix \ref{app:Px}.  Equations \ref{eq:cov}, \ref{eq:Px} and \ref{eq:sig} constitute our analytic model for the cross-correlation.

\section{Comparison to simulations}\label{sec:sims}

The analytic form above makes several simplifying assumptions (e.g.\ neglect of the survey window function and higher-order contributions to the trispectrum) that are wrong in detail, and we thus wish to check how well it works by comparison to simulations that include all of these effects.  To this end we implemented this analytic expression and here we compare it to covariance matrices obtained from mock catalogs (from simulations) generated in support of the DESI project \cite{Chuang2015}.

\subsection{Cubic geometry}

We begin by testing our approximation within a cubic box geometry. We use as data vectors the pre-reconstruction $P_{\ell}(k)$ and post-reconstruction $\xi_{\ell}(s)$ spectra from the Abacus cubic LRG mocks (mean of 25 realizations) that were used in ref.~\cite{KP5s2-Maus} for theory validation tests for the DESI DR1 full-shape analysis \cite{DESI2024.V.KP5}. The cubic mocks were produced at a redshift snapshot of $z=0.8$. Since these simulation boxes have periodic boundary conditions, their window function is trivial. The corresponding numerical covariances are computed from 1000 EZmocks \cite{Chuang2015} and correspond to an $8\,(h^{-1}\mathrm{Gpc})^3$ volume. 

In our analytic approximation for the Pre$\times$Post covariance (eqs.~\ref{eq:cov}-\ref{eq:sig}) we use $V=8\,(h^{-1}\mathrm{Gpc})^3$ corresponding to the volume of the mocks. We use the same Abacus input cosmology as was used in the mocks to generate the linear power spectrum $P_{\rm lin}(k)$ using the \texttt{CLASS} Boltzmann solver~\cite{CLASS} and likewise compute the fiducial growth rate $f(z=0.8) = 0.838$. For the Eulerian linear galaxy bias we use $b=2.03$ obtained from best fits to these LRG mocks in ref.~\cite{KP5s2-Maus}. We find very good agreement between the analytic approximation and the simulation based covariance in each $P_{\ell_1}(k_i)\times \xi_{\ell_2}(r_j)$ slice --- in the next section, we show this explicitly for the more demanding case of a non-trivial observational window.

\begin{figure}
    \centering
    \begin{subfigure}{0.48\textwidth}
        \centering
        \includegraphics[width=\textwidth]{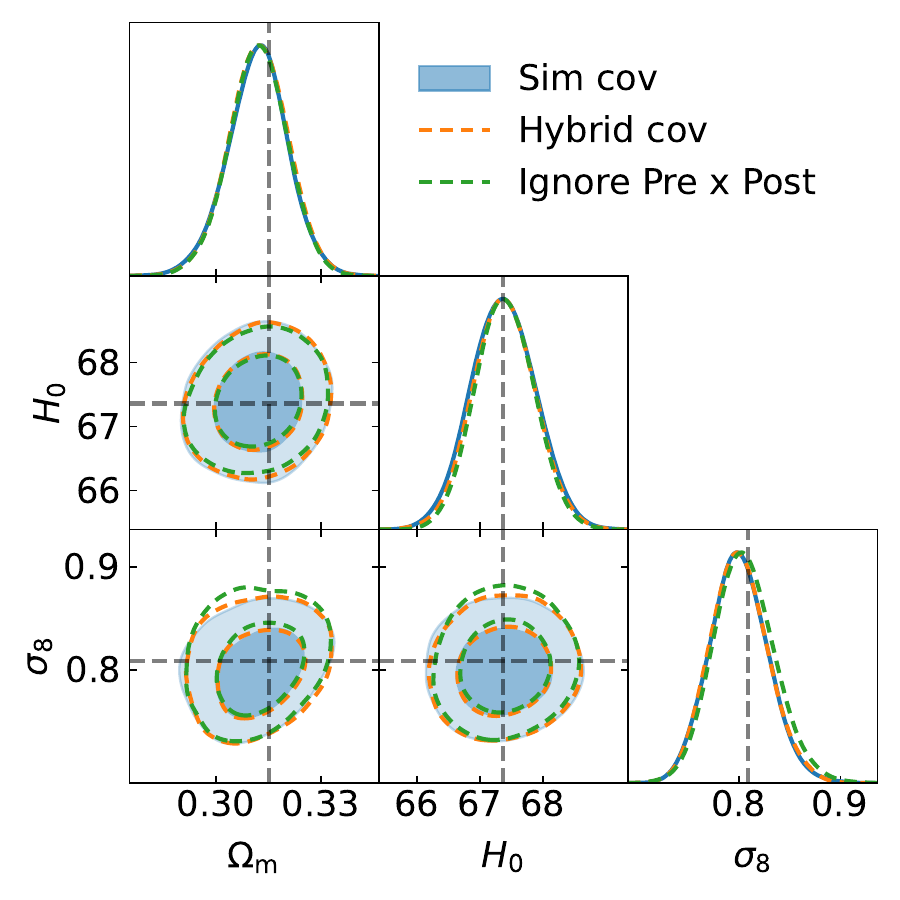}
        \caption{Cubic geometry}
        \label{fig:contours_cubic}
    \end{subfigure}
    \hfill
    \begin{subfigure}{0.48\textwidth}
        \centering
        \includegraphics[width=\textwidth]{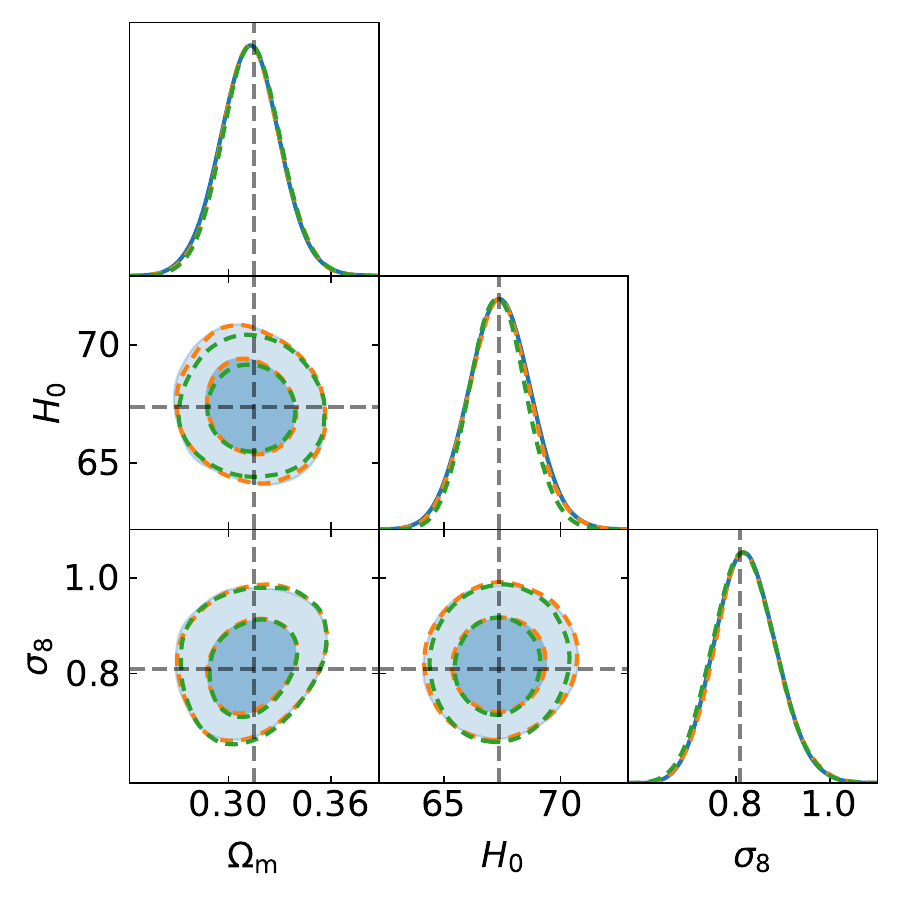}
        \caption{Cut-sky geometry }
        \label{fig:contours_cutsky}
    \end{subfigure}
    \caption{Cosmological parameters inferred from mocks using a covariance matrix where the off-diagonal  $P_{\ell_1}(k)-\xi_{\ell_2}(r)$ block is determined analytically (orange), numerically (blue), or it is ignored altogether (green). Dashed lines show the true parameter values underlying the mocks.  In the left panel, the simulation box is cubic and periodic, whereas in the right the geometry simulates the DESI footprint and lightcone structure and includes also a $\theta$-cut.}
    \label{fig:combined}
\end{figure}

To test the impact of approximate covariance treatments on parameter constraints, we create two new covariances from the simulation based covariance: one hybrid version with the off-diagonal $P_{\ell_1}(k_i)\times \xi_{\ell_2}(r_j)$ blocks replaced with our analytic approximation\footnote{When building hybrid covariances in this way, we find it necessary to null eigenvectors of the composite matrix that have negative eigenvalues (likely due to noise) in order to ensure that it remains positive-definite. The resulting matrix can be thought of as the positive-definite matrix that is closest to the original one under the Frobenius norm. Concretely, we have to null just two negative eigenvalues whose impact is largely negligible -- their amplitudes are less than 1\% and 0.1\% of the largest eigenvalue. Along the diagonal the structure associated with these negative eigenvalues never accounts for more than a 10\% change to individual entries, and typical per-bin variations are much smaller than that. More generally, these negative eigenvectors act to decrease the amplitude of the $\xi_0(r_i)\times \xi_0(r_j)$ and $\xi_2(r_i)\times \xi_2(r_j)$ blocks, increase their cross-covariance, and imprint slight tilts in $k$ to the $P_a(k_i)\times \xi_b(r_j)$ blocks about a pivot scale of $k\approx 0.05\,h \rm{Mpc}^{-1}$ --- the tilt is red for terms involving $\xi_2$, and blue for $\xi_0$.}, and another with those blocks set to zero (i.e.\ ignoring the Pre$\times$Post covariance entirely). The constraints, shown in fig.~\ref{fig:contours_cubic}, are very similar in all three cases, with the hybrid case matching the fully simulation-based covariance case almost exactly. The marginalized constraints on individual parameters are also quoted in table~\ref{tab:constraints}, where the analytic treatment of the pre-post covariance gives error bars that never deviate from the fully simulation-based ones by more than 4\% for any individual parameter, while ignoring this covariance results in deviations of up to 10\%.
\begin{table*}
\centering
\caption{Mean and standard deviation of the marginalized posterior on cosmological parameters for different simulation geometries and treatments of the covariance between pre-reconstruction $P_{\ell}(k)$ and post-reconstruction $\xi_{\ell}(s)$.}
\label{tab:constraints}
\resizebox{\textwidth}{!}{%
\begin{tabular}{l | ccc | ccc | ccc}
\hline\hline
 & \multicolumn{3}{c|}{Cubic geometry} & \multicolumn{3}{c|}{Cutsky geometry} & \multicolumn{3}{c}{Cutsky geometry + window rotation} \\
& Simulated & Analytic & Ignored & Simulated & Analytic & Ignored & Simulated & Analytic & Ignored \\
\hline
$H_0$ 
    & $67.37 \pm 0.51$ 
    & $67.37 \pm 0.49$ 
    & $67.39 \pm 0.47$ 
    & $67.40 \pm 1.33$ 
    & $67.38 \pm 1.31$ 
    & $67.31 \pm 1.20$ 
    & $67.64 \pm 1.59$ 
    & $67.56 \pm 1.59$ 
    & $67.42 \pm 1.45$ \\[8pt]
$100\,\Omega_m$ 
    & $31.21 \pm 0.84$ 
    & $31.21 \pm 0.83$ 
    & $31.20 \pm 0.82$ 
    & $31.25 \pm 1.74$ 
    & $31.33 \pm 1.71$ 
    & $31.36 \pm 1.69$ 
    & $31.29 \pm 2.11$ 
    & $31.36 \pm 2.04$ 
    & $31.39 \pm 2.01$ \\[8pt]
$100\,\sigma_8$ 
    & $80.0 \pm 2.8$ 
    & $79.9 \pm 2.8$ 
    & $80.3 \pm 3.0$ 
    & $81.8 \pm 6.5$ 
    & $81.9 \pm 6.5$ 
    & $81.8 \pm 6.7$ 
    & $82.1 \pm 7.8$ 
    & $81.6 \pm 7.9$ 
    & $81.6 \pm 7.9$ \\
\hline\hline
\end{tabular}%
}
\end{table*}

\subsection{Non-trivial geometry and $\theta$-cut}

When analyzing real data in Fourier space, one must account for the mode-coupling induced by observational effects -- collectively, these are known as the survey's window function. Theoretical predictions must be convolved with this window before they can be compared to data, and the covariance must in principle also account for it. 

Two crucial contributors to the window function of galaxy redshift surveys are the geometry of the survey's footprint and the sample incompleteness related to fiber assignment. The latter stems from the fact that objects selected for spectroscopic follow-up compete for fiber assignment, such that it may take several telescope visits to a given region of sky to obtain a fair sample of the population. In the meantime, mitigation schemes must be implemented to avoid obtaining biased inferences. One such scheme, adopted in the DESI DR1 full-shape analysis \cite{DESI2024.V.KP5}, is the $\theta$-cut \cite{KP3s5-Pinon} method.  In this approach, pairs of galaxies are removed from the power spectrum estimators if their angular separation is below the scale of fiber collisions. This requires modifying the window function accordingly. 

Incorporating the effect of the fiber collisions and the survey geometry into an analytic covariance is not a trivial task, although it is possible that existing pathways could be repurposed to this end~\cite{Li2019}. However, the results above show that the $P_{\ell_1}^{\rm pre}\times\xi_{\ell_2}^{\rm post}$ covariance is small, suggesting that any corrections due to the window function may well be negligible. Let us now validate this assumption.

Specifically, we now check that the window function can be safely neglected in our analytic $P_{\ell_1}^{\rm pre}\times\xi_{\ell_2}^{\rm post}$ blocks of the joint covariance, so long as they are included in the $P^{\rm pre}_{\ell_1}\times P^{\rm pre}_{\ell_2}$ blocks. We do this by once again comparing our hybrid covariance to one derived from mocks, only that now we include non-trivial observational effects in the latter by applying the DESI DR1 footprint and $\theta$-cut to the \texttt{EZmocks} before measuring clustering statistics.  We will refer to these as the `cut-sky' mocks. Certain applications of the $\theta$-cut method apply a rotation of the data vector, window matrix and covariance matrix designed to make the transformed window more compact in Fourier space; in appendix~\ref{sec:window_rot} we show that our vanilla analytic result remains a good approximation even after rotation.

\begin{figure}
    \centering
    \includegraphics[width=0.99
    \textwidth]{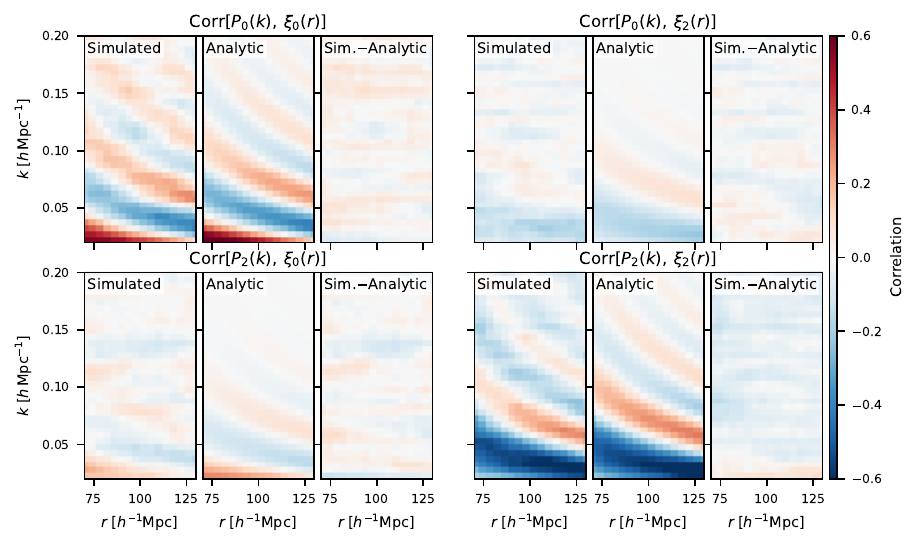}
    \caption{Correlation between multipoles of the pre-reconstruction galaxy power spectrum and post-reconstruction correlation function. For each one of the four sub-blocks of this correlation matrix, we show the matrix estimated from simulations on the left, the analytic prediction in the center, and the difference between the two estimates on the right, which reveal only small differences in all cases. The catalogs used to produce this simulated covariance have a cutsky geometry and a $\theta$-cut, but no window rotation (we have checked that the agreement is just as good when the latter is included).
    }
\label{fig:correlation_2D}
\end{figure}

\begin{figure}
    \centering
    \includegraphics[width=0.99
    \textwidth]{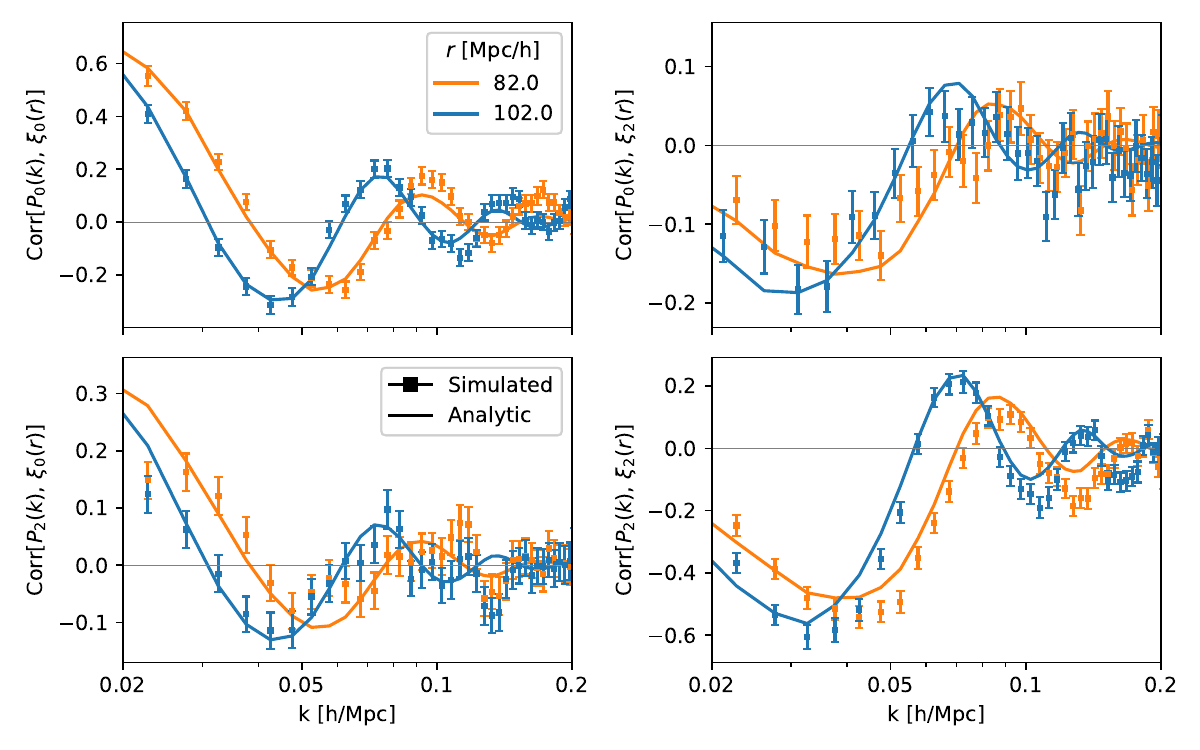}
    \caption{Cross-correlation matrix between $P_{\ell_1}(k)$ and $\xi_{\ell_2}(r)$ calculated from our analytic formula (dashed lines) or from simulations (points) as a function of $k$ for two different values of the pair separation $r$ (colors).  In order to focus on the accuracy of the pre-post covariance, we divide by the same simulation-derived variances when calculating the correlation matrix. The errors on the simulated points are estimated by leave-one-out jackknife.  The simulations have cut-sky geometry and a $\theta$-cut, but no window rotation is applied. In the upper right panel we added a small leftward horizontal shift ($\Delta k=-0.0015 \ihmpc$) to the blue data points in order to reduce overlap with the blue errorbars. 
    }
\label{fig:off_diag_blocks_cutsky}
\end{figure}

We show in fig.~\ref{fig:correlation_2D} the correlation structure of the $P_{\ell_1}^{\rm pre}\times\xi_{\ell_2}^{\rm post}$ block, comparing simulated and analytic results. Then, in fig.~\ref{fig:off_diag_blocks_cutsky}, we zoom in on the $k$-dependence of the correlation at two fixed $r$-bins, $r=82$ and $102\,h^{-1}$Mpc. The pre-post covariance drops faster with $k$ than the variances of the individual blocks do, and by $k\approx 0.1 \,h\,\rm{Mpc}^{-1}$ the correlation between $P_{\ell_1}(k)$ and $\xi_{\ell_2}(r)$ is already down to 20\%. Our analytic model is able to predict this, and we find good agreement with the simulation-based result across a wide range of scales despite ignoring window effects and $\theta$-cut. In our model we use $V=2.87\,(h^{-1}\mathrm{Gpc})^3$ for the effective volume\footnote{We approximate the effective volume as $[\int n^2(\mathbf{x})w^2(\mathbf{x})]^2/\int n^4(\mathbf{x})w^4(\mathbf{x})$, with mean density $n(\mathbf{x})$ and window/weights $w(\mathbf{x})$, which can be evaluated with \texttt{TheCov} as \texttt{geometry.I('22')**2/geometry.I('44')}. This factor provides the correct scaling in the limit that the window is much narrower than the bin width in Fourier space (see e.g. Ref.~\cite{Li2019} for a recent discussion).}, as measured from the randoms of the mocks, and an Eulerian bias of $b=2.05$, obtained from fitting the DESI DR1 LRG ($0.6\leq z \leq 0.8$) sample. For the shotnoise term, SN, in eq.~\ref{eq:Px} we use $0.6\times(1/\bar{n})$ with $1/\bar{n} = 5000\,h^{-3}\mathrm{Mpc}^3$ because ref.~\cite{Kokron22} found that the stochastic contributions to the power spectra for LRGs such as those considered in this work have $\bar{n}P_{\rm err}\approx 0.6$. The errorbars on the simulated covariance bins are estimated using a jackknife method wherein we compute the covariance from 999 of the 1000 EZmock spectra, leaving out a different realization each time. The standard error is then the square root of $(N-1)\times$ the average squared deviation from the jackknife mean. 

To check the impact of any differences at the parameter inference level we use synthetic data computed from our $P_{\ell_1}^{\rm pre}+\xi_{\ell_2}^{\rm post}$ theoretical pipeline with the cosmology fixed to the fiducial Abacus cosmology and nuisance parameters conditioned on the DESI DR1 data. This theoretical model is then convolved with the DESI DR1 window function to form a synthetic $P^{\rm pre}(k)$ with the same $k$-binning as used in ref.~\cite{DESI2024.V.KP5}. As in the cubic case we fit the data using either the completely simulation-based covariance, a hybrid with $P_{\ell_1}^{\rm pre}\times\xi_{\ell_2}^{\rm post}$ replaced with the analytic approximation, and one with the $P_{\ell_1}^{\rm pre}\times\xi_{\ell_2}^{\rm post}$ contribution set to zero.

Fig.~\ref{fig:contours_cutsky} and table~\ref{tab:constraints} present the comparison of the parameter constraints.  Despite the approximations we have made in our analytic calculation, we still find excellent agreement in constraints between the fully simulation-based case and when using the hybrid covariance.  Ignoring the  $P_{\ell_1}^{\rm pre}\times \xi_{\ell_2}^{\rm post}$ covariance block entirely performs less well, but is still very close to the full simulation-based constraints.


\section{Discussion and conclusion}\label{sec:conclusions}

It is now routine for galaxy redshift survey collaborations to analyze the combination of post-reconstruction correlation functions and pre-reconstruction power spectra.  The BAO signal in the former provides tight constraints on the background cosmology through the distance-redshift relation while the detailed shape of the latter provides constraints on the cosmological parameters and growth of structure.  The combination has proven particularly powerful.  However these two probes are not uncorrelated as they both trace the same underlying matter field.  In fact, in linear theory they would be completely degenerate if measured over the full range of scales.

In reality the presence of non-linearity, shot noise and finite scale ranges damp this correlation.  Thus in practice the degree of correlation is quite small.  The smallness of this correlation suggests that even an approximation to it may be sufficient for data analysis.

In this paper we have provided a simple analytic model for the covariance between the two measurements (equations \ref{eq:cov}, \ref{eq:Px} and \ref{eq:sig}).  Using mock catalogs from DESI DR1 we have demonstrated that the covariance has a small impact on cosmological parameter inference from the joint datatset and that inclusion of the covariance through our analytic model returns essentially identical constraints to those with the mock-determined covariance including all relevant survey non-idealities.

The analytic approximation provides an easy route to fitting combined BAO and full-shape statistics that we have found useful in our work.  Looking ahead, having a reasonable, noiseless analytic approximation may also prove useful for denoising numerical estimates that are based on Monte-Carlo evaluations of a large number of mock catalogs or provide a means for `updating' covariance matrices under small changes in the fiducial cosmology.  The model can be used as a stop-gap measure to aid in code development while mock catalog generation is ongoing.  When combined with other efforts to compute covariance matrices of pre- and post-reconstruction statistics it provides a pathway towards fully analytic treatments of covariances that could even remove the need for running very large numbers of low-fidelity, large-volume mocks.  This would allow computational resources to be reduced or focused elsewhere.  Along these lines, the small number of parameters involved may make it possible to use a data-driven approach (e.g.~jackknife) for tuning the bias, shotnoise and damping parameters (potentially including theory-inspired priors).

We make our code publicly available.\footnote{
\texttt{PrePostCov}: \href{https://github.com/abaleato/PrePostCov/tree/main}{https://github.com/abaleato/PrePostCov}}

\section*{Acknowledgments}
We thank Naonori Sugiyama for illuminating discussions on the pre- and post-reconstruction power spectrum covariance, particularly the role of long-wavelength displacements in their decorrelation. We would like to thank Otávio Alves and Misha Rashkovetskyi for help with \texttt{TheCov} and \texttt{RascalC}, respectively.  MM and MW were supported by the DOE.  This research was supported in part by grant NSF PHY-2309135 to the Kavli Institute for Theoretical Physics (KITP). Support for this work was provided by NASA through
the NASA Hubble Fellowship grant HST-HF2-51572.001
awarded by the Space Telescope Science Institute, which
is operated by the Association of Universities for Research in Astronomy, Inc., for NASA, under contract
NAS5-26555.
This work made use of the Cobaya analysis code \cite{Torrado21,CobayaSoftware} and the resulting MCMC chains were analyzed and plotted using GetDist \cite{Lewis19}.
This research used resources of the National Energy Research Scientific Computing Center (NERSC), a Department of Energy User Facility.

\appendix

\section{Non-trivial geometry and $\theta$-cut with window rotation}
\label{sec:window_rot}

\begin{figure}
    \centering
    \includegraphics[width=0.99
    \textwidth]{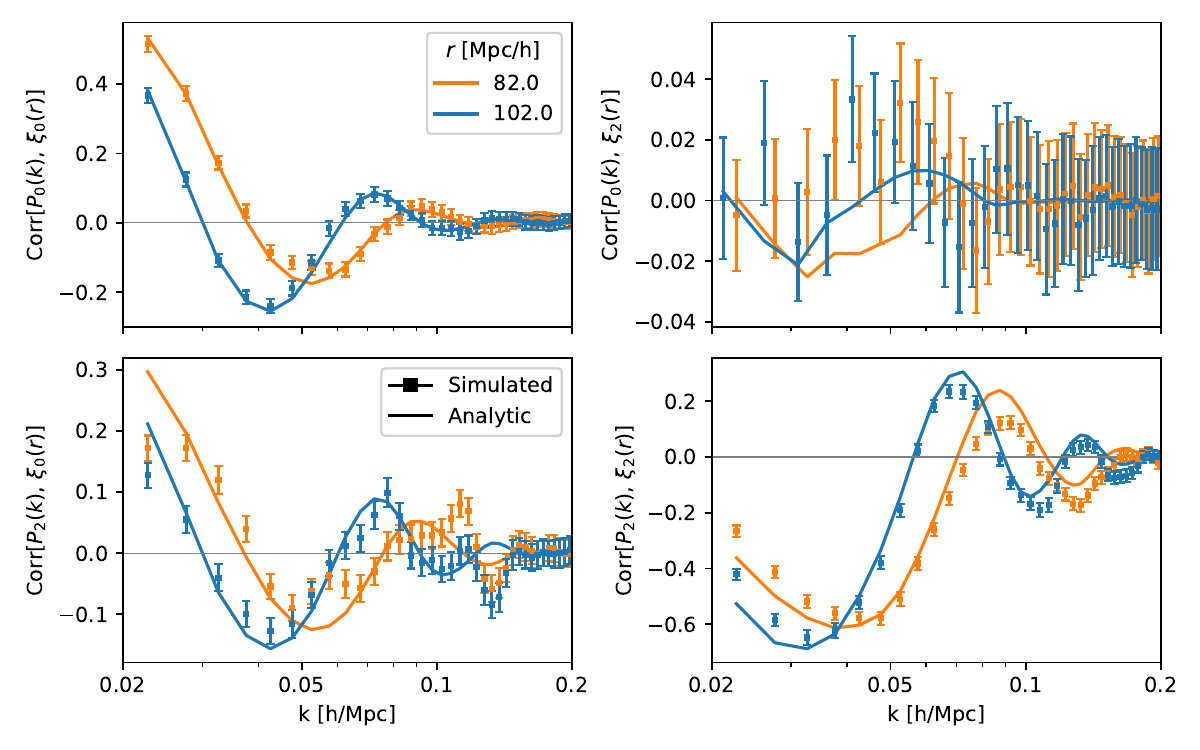}
    \caption{Same as figure~\ref{fig:off_diag_blocks_cutsky}, but with window rotation following the $\theta$-cut.}
\label{fig:off_diag_blocks_cutsky_rot}
\end{figure}

\begin{figure}
    \centering
    \includegraphics[width=0.55
    \textwidth]{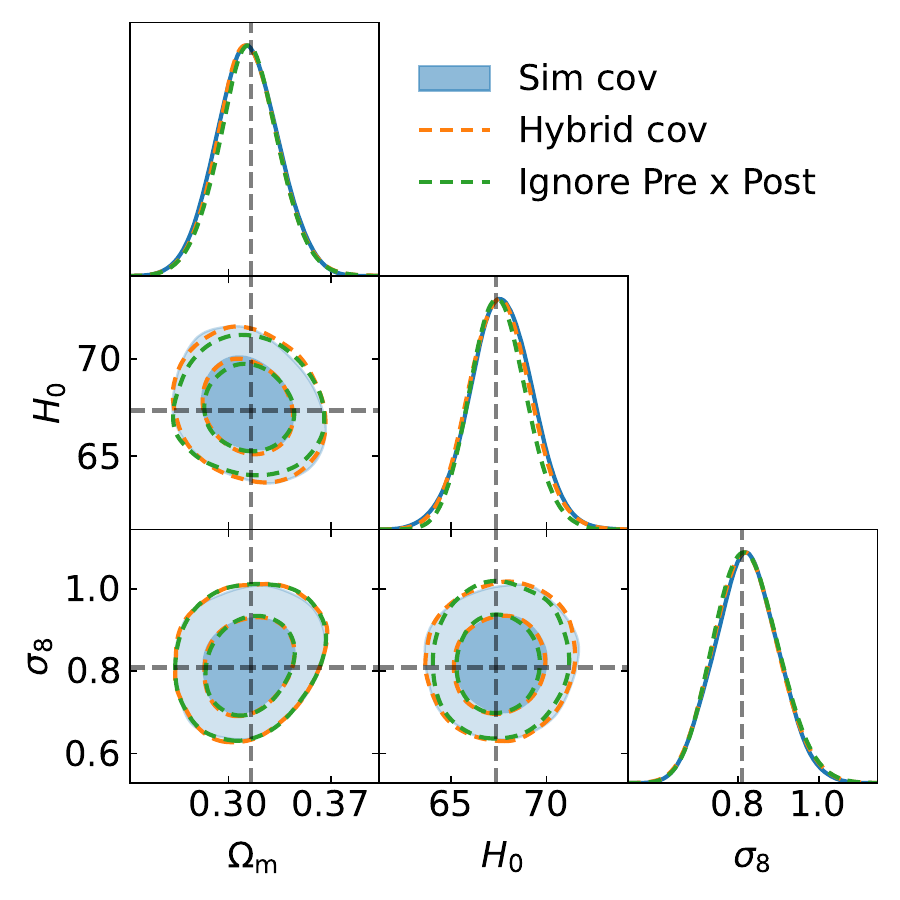}
    \caption{Same as Fig.~\ref{fig:combined} but with the rotation applied to diagonalize the window after the $\theta$-cut has been applied. }
\label{fig:contour_cutsky_rot}
\end{figure}

Finally, we note that in the official DESI DR1 analysis \cite{DESI2024.V.KP5} a rotation was applied consistently to the data vector, window and covariance matrix in order to limit the extent of the window in Fourier space. This ensures that $k$-modes used in the analysis are less affected by high-$k$ modes outside of the range of validity of the theory \cite{KP3s5-Pinon}. The most salient impact of this rotation on the covariance matrix is that, as shown in Fig.~\ref{fig:off_diag_blocks_cutsky_rot}, the signal in the $P_0\times \xi_2$ block is suppressed. After applying the rotation matrix to our analytic result, the model is consistent with the numerical computation. As in the previous cases we test our model on the parameter inference level by replacing the $P_{\ell}^{\rm pre}\times \xi_{\ell}^{\rm post}$ blocks of a mock-based covariance and comparing constraints to those obtained when using the full mock-based covariance as well as when ignoring the $P_{\ell}^{\rm pre}\times \xi_{\ell}^{\rm post}$ contributions. We show the results in Fig.~\ref{fig:contour_cutsky_rot} and table~\ref{tab:constraints}. We find that the posteriors when using the hybrid covariance very closely matches those from the numerical covariance. Meanwhile, when ignoring $P_{\ell}^{\rm pre}\times \xi_{\ell}^{\rm post}$ contributions the constraints in the $H_0-\Omega_{\rm m}$ plane narrow slightly, suggesting that ignoring these blocks of the covariance causes an artificial underestimation of error bars when reporting constraints on these parameters.

\section{Model for the cross-power spectrum}
\label{app:Px}

In the main text we presented a perturbative model for $P_\times$, the cross-power between the pre- and post-reconstruction power spectra.  In this appendix we provide further context on this result and its relationship to the broadening of the BAO peak.  Since we lose nothing essential by neglecting redshift-space distortions, we shall work in real space in what follows.

To lowest order the large-scale displacement of a reconstructed galaxy from its initial position can be described in terms of a high-pass filtered linear (Zeldovich) displacement $\Psi^{(1)}_{\rm rec}(\bk) = [1 - W(k)]\Psi^{(1)}(\bk)$. In this approximation the pairwise displacement between a reconstructed and un-reconstructed galaxy $\mathbf{\Delta}(\bq_1,\bq_2) = \mathbf{\Psi}_{\rm rec}(\bq_1) - \mathbf{\Psi}(\bq_2)$ has covariance
\begin{equation}
    A_{ij}(\bq_1 - \bq_2) = \langle \Delta_i \Delta_j \rangle = \langle \Psi_{\rm rec, i} \Psi_{\rm rec, j} \rangle + \langle \Psi_{i} \Psi_{j} \rangle - 2 \langle \Psi_{\rm rec, i}(\bq_1) \Psi_{j}(\bq_2) \rangle.
\end{equation}
In the small- and large-separation limits
\begin{equation}
    A_{ij}(\mathbf{q}\to 0) = \delta_{ij} \Sigma^2_{0}
    \qquad , \qquad
    A_{ij}(\bq \rightarrow \infty)= \delta_{ij} \left\{\Sigma^2_0 + \frac{2}{3}\int \frac{dk}{2\pi^2} [1 - W] P(k) \right\}
\end{equation}
where $\Sigma^2_{0} = \sigma^2(\mu=0)$. The limit of small-scale clustering ($\xi(r) \sim \delta_D(r)$), which is the pertinent one to reproduce the broadband of the power spectrum most relevant for the covariance, corresponds to the $\mathbf{q}\to 0$ limit. In this regime, the effect on the clustering is given by \cite{CLPT}
\begin{equation}
    P_{\rm small-scale}(k) \simeq \int d^3\bq \ e^{i\bk \cdot \bq - \frac{1}{2} k_i k_j A_{ij}(\bq)}\ A \delta_{D}(\bq) = A e^{-\frac{1}{2} k^2 \Sigma_0} \quad ,
\end{equation}
i.e.\ the damping is due to the variance of the pairwise displacement evaluated at a point. This limit is exact for the shot noise, and this is also the approximation used in the main body of the paper.

On the other hand, this approximation over-predicts the amplitude of the BAO in the cross, which instead depends on the variance evaluated at $q = |\bq| = r_d$ and $\hat{k} = \hat{q}$,  where \cite{CVW24}
\begin{equation}
    k_i k_j A_{ij}(\bq) = k^2 \left\{ \Sigma^2_0 + \frac{2}{3} \int \frac{dk}{2\pi^2} [1 - W(k)] \left[ 1 - j_0(k r_d) + 2 j_2(k r_d) \right] P_{\rm lin}(k) \right\} \ \ .
\end{equation}
Using \texttt{velocileptors} \cite{Chen20,Chen21} we have numerically verified that the damping scale described by these two limits accurately captures the broadband and BAO features in a toy model where the initial distribution of galaxies is given by a linear bias, $\delta_g(\bq) = b_1 \delta_{\rm lin}(\bq)$.

\bibliography{main}
\bibliographystyle{jhep}

\end{document}